\documentclass[a4paper]{article}

\usepackage[english]{babel}
\usepackage[utf8x]{inputenc}
\usepackage[T1]{fontenc}

\usepackage[a4paper,top=3cm,bottom=2cm,left=3cm,right=3cm,marginparwidth=1.75cm]{geometry}

\usepackage{amsmath}
\usepackage{graphicx}
\usepackage[colorinlistoftodos]{todonotes}
\usepackage[colorlinks=true, allcolors=blue]{hyperref}

\usepackage{lineno}
\pdfoutput=1 
\usepackage{multirow}
\usepackage{caption}
\usepackage{subcaption}
\usepackage{eqnarray}
\usepackage{amssymb}
\usepackage{url}

\usepackage{authblk}

\title{{Input from the CENF-ND Forum to the 2020 Update of \\ the European Strategy for Particle Physics}\\
\vspace{0.2cm}
\large{
Research and Development for Near Detector Systems Towards Long Term Evolution of Ultra-precise Long-baseline Neutrino Experiments.
}
}

\author[6]{L.~Alvarez Ruso}
\author[12]{J.~Asaadi}
\author[2]{S.~Bolognesi}
\author[3]{S.~Bordoni\thanks{stefania.bordoni@cern.ch}}
\author[3]{A.~de Roeck}
\author[1]{M.V.~Diwan}
\author[5]{T.~Lux}
\author[9]{D.~Meloni}
\author[3]{M.~Nessi}
\author[10,11]{B.~Popov}
\author[7]{E.~Radicioni}
\author[3,8]{P.~Sala\thanks{paola.sala@cern.ch}}
\author[4]{F.~Sanchez}
\author[3]{L.~H.~Whitehead}

\affil[1]{Physics Department, Brookhaven National Laboratory, Upton, NY, USA}
\affil[2]{CEA-Saclay, Paris, France}
\affil[3]{CERN, European Organization for Nuclear Research, Geneva, Switzerland}
\affil[4]{University of Geneva, Geneva, Switzerland}
\affil[5]{IFAE, Barcelona, Spain}
\affil[6]{Instituto de F\'isica Corpuscular (IFIC), Valencia, Spain}
\affil[7]{INFN Sezione di Bari, Bari, Italy}
\affil[8]{INFN Sezione di Milano Bicocca, Milano, Italy}
\affil[9]{Dipartimento di Matematica e Fisica, Universit\`{a} degli Studi Roma Tre, Roma, Italy}
\affil[10]{Joint Institute for Nuclear Research, Dubna, Russia}
\affil[11]{LPNHE, Paris, France}
\affil[12]{Univerity of Texas at Arlington, Arlington, TX, USA}

\begin{document}

\maketitle
\begin{center}
\large{(On behalf of the CERN CENF-ND Forum)}
\end{center}
\begin{abstract}
With the discovery of non-zero value of $\theta_{13}$ mixing angle, the next generation of 
long-baseline neutrino (LBN) experiments offers the possibility of obtaining 
statistically significant samples of muon and electron neutrinos and 
anti-neutrinos with large oscillation effects.
In this document we intend to highlight the importance of Near Detector facilities in LBN experiments to both constrain the systematic uncertainties affecting oscillation analyses but also to perform, thanks to their close location, measurements of broad benefit for LBN physics goals. A strong European contribution to these efforts is possible.
\end{abstract}

\flushbottom
\clearpage

\section{Introduction}

The next generation of long-baseline neutrino (LBN) experiments -- LBNF/DUNE in the US and T2HyperK (T2HK) in Japan -- will make precise measurements of neutrino oscillation parameters with a focus on the determination of the CP violating phase $\delta_{CP}$ in the 3-neutrino PMNS oscillation formalism.  Furthermore, these experiments will collect sufficient statistics in measured samples of muon and electron neutrinos that a precision fit could reveal unexpected physics beyond the 3-neutrino PMNS framework.  


The scientific performance of both T2HyperK and LBNF/DUNE projects depends crucially on the ability to precisely predict the spectra in the far detectors, a liquid argon TPC for DUNE and a very large water Cherenkov detector for Hyper-K.  Both DUNE and Hyper-K experiments require very large and long term investments from the international particle physics communities. The ultimate statistical precision from these experiments is expected to reach a few percent level after a decade of running with high power accelerator neutrino beams. Along with the accumulation of statistics, a step-by-step program of improvements to the systematic errors is also needed.  European physicists using the facilities at CERN and other European laboratories can play a leading role in the program of precision neutrino physics with a long term focus on sub-percent level of understanding of accelerator neutrino beams and neutrino-nucleus interactions in the near and far detectors.  \\

In all long-baseline accelerator neutrino experiments, the best results are obtained by measuring the event rate and spectra in a near detector placed at the accelerator laboratory site before the beam traverses a sufficient distance to be affected by oscillations. The results from the near detector measurements are extrapolated to the far detector using the knowledge of the geometry, the beam simulation, and differences in the near and far detector efficiencies.  The experience from the recent experiments (K2K,  MINOS,  T2K, and NO$\nu$A) has steadily increased the precision in long-baseline experiment from  >10\% to few-percent level over the last two decades.  This path to high precision needs to be continued for the success of LBNF/DUNE and T2HK, in addition to the technological improvements in the field of accelerator neutrino beams.  We have systematically examined these past approaches and have concluded that a program of steady improvements to the near detector site, the near detector technology, the neutrino beam instrumentation, and simulations guided by data, is needed to achieve the ultimate goals.  This document intends to be a coherent plan for executing this program with European and CERN based leadership.  

Our plan derives from the material and discussions generated as part of the CERN Neutrino Platform activities on near detectors, the so called "CENF-Near Detector Forum", which is presented in Section~\ref{sec:CENFND}. The materials included a survey of current techniques as well as ideas and proposals for future work.

We have classified the future program of activities in three categories:

\begin{itemize}
\item   Near detector technologies and their impact on far detector predictions.
\item	Considerations of beam and near detector configuration for the next generation of experiments.
\item   Ancillary measurements and improvements of the current Monte Carlo modelling of neutrino interactions.
\end{itemize}

In the rest of this report we briefly expand on specific items regarding each of these categories and their impact on far detector predictions and precision oscillation measurements. 

In Section~\ref{sec:NDtech} we review the importance of near detectors for LBN analyses, discussing the measurements which can be achieved at the near site and providing some practical considerations about how these options could be implemented.  In Section~\ref{sec:ND} we discuss the beam and the planned near detector site along with various physical and economical constraints for LBNF/DUNE and T2HK. 
We expand on the possibilities proposed in the CENF working group and possible implementation.     In Section~\ref{sec:AncMeas} we provide information on ancillary measurements of neutrino cross sections and meson production cross sections and benefits of these to the LBNF/DUNE and Hyper-K physics programs.  
In Section~\ref{sec:Generators} we discuss the need of improvements in the current Monte Carlo neutrino event generators and how this can be addressed.  
Finally in Section~\ref{sec:Conclusions} we derive our conclusions.

\section{The CENF Near Detector Forum}
\label{sec:CENFND}

The CENF-Near Detector forum (CENF-ND)~\cite{CENFND} is a CERN Neutrino Platform activity started in 2017. The aim is to strengthen the European effort on the current and next generation of accelerator-based Long Baseline Neutrino experiments as well as to expand the Neutrino Platform connections to American and Japanese collaborators. 
As explicit in its name, the focus of the forum is on the LBN Near Detectors technologies and Near Detector facilities which have been demonstrated by the current experiments like T2K, NO$\nu$A, MINOS+ to be crucial to achieve the precision measurements foreseen by the LBN experiments physics goals for the next few years.

\section{The role of Near Detectors}
\label{sec:NDtech}

Neutrino beams used for LBN experiments are generally characterized by broad energy spectra. Off-axis techniques can be used, as done by T2K and NO$\nu$A, to narrow the spectrum; nevertheless neutrino mono-energetic beams cannot be produced by standard accelerator facilities. 
Near detectors play a unique and crucial role for LBN experiments allowing for a number of measurements: monitor the incoming beam in terms of rate, characterize its composition before any oscillation and reconstruct the initial neutrino energy from the interaction products. This last point also provides the possibility to constrain common uncertainties related to the detector response if near and far sites base their measurements on a common technology. 
To quantify the impact of near detectors on oscillation analyses we cite as example the recent results from the T2K experiment. Thanks to the constraints coming from the near site, the uncertainties affecting the number of interactions at the far site are reduced by about a factor of two: for the $\nu_\mu$ ($\nu_e$) one-ring event selections in neutrino mode from 11\% (12\%) to 4\% (5\%) \cite{T2K_NDconstrain}. 
Near detectors, thanks to their location close to the beam production site and therefore the large statistics available, offer also the possibility to study neutrino interactions by measuring their cross-section as a function of different observables such as, for example, the initial momentum, the incident angle, the transferred four momentum squared Q$^2$ or other derived quantities which can give access to nuclear effects in neutrino interactions. Furthermore such detectors can play a leading role in searches for non-standard processes~\cite{PONDD}. 

The choice of the optimal near detector technologies depends on the beam characteristics as well as on the technology chosen for the far detector site. There are two main strategies applied by the current experiments: the choice of the same technology for both near and far site (e.g. NO$\nu$A) or a general purpose detector at the near site allowing also for an independent and rich physics program (e.g. T2K). For the future generation of LBN experiments the choice of a multiple technology detector at the near site seems to be more appropriate. This choice is driven by the need of improving the current understanding of neutrino interactions and the related systematic uncertainties affecting the far detector measurements.   \\

Innovative ideas like ENUBET~\cite{ENUBET1}-- a proposal for an electron-neutrino beam from the three body semi-leptonic decay of kaons -- and NuSTORM\cite{nustorm} a muon storage ring which would allow very well know fluxes of $\nu_\mu , \nu_e, \overline{\nu}_\mu ,\overline{\nu}_e $-- have been also discussed at the CENF-ND Forum.
Such programs are of deep interest for the community since they would provide (anti-) neutrino fluxes at very few percent level and both electron and muon cross-section measurements with unprecedented precision. However such projects are on a different size with respect to the actions suggested in this document.
Another project which has been discussed is the ESSnuSB design study which considers to use the 5 MW European Spallation Source linear accelerator in Sweden as proton driver. The high intensity neutrino beam resulting from the (5 MW) 2~GeV proton beam will allow the far neutrino detector to be placed at the second neutrino maximum where the signal is enhanced about 3 times compared to the first maximum~\cite{ESSnuSB}.  Near detector options are being studied with the goal to optimise the technology towards low momentum neutrinos therefore some of the items discussed in the document can be of interest also for this project.

\section{Near Detectors Location and Configuration}
\label{sec:ND}

After the decisions on the far detector and the neutrino beam energy, 
the most critical points to be defined in the long baseline experiments are the distance to the near detector, and the number and type of near detectors. 
For an accelerator neutrino beam, the source is both extended over the length of the meson decay tunnel, and the beam collimation varies with energy; 
therefore, to first order, the average  source location appears to vary with the energy spectrum creating a strong near/far asymmetry in the energy spectrum.  The problem is made even more difficult because the nature of the near detector is unlikely to be identical to the technology at the far site due to practical difficulties of hall size and interaction pileup rate.  

Three constraints have determined the locations of the near detector hall in the LBNF/DUNE and in the T2HK programs.  First, the near detector has to be far enough to avoid the flash of muons from the beam dump at the end of the decay tunnel.  
Second, the near detector should be far enough to minimize the near/far neutrino spectra differences.  Third,  the detector needs to  remain on the site of the accelerator laboratory to minimize cost and difficulty of construction.   

The spectrum of neutrino events in the far detector of long baseline experiments results from a convolution of many effects. 
Briefly, the probability density of mesons produced in the target and focused by the horns gets convoluted with the probability of neutrino interactions in the near and far detectors and further with the efficiency of reconstruction, particle identification, and energy resolution.  The final un-oscillated spectrum at the far site must be determined by careful evaluation of near site observations and the joint correlations between the near and far detectors.   
Consequently, there is a long list of imperfect correlations due to the evolution of the energy spectrum as a function of distance from the target, cross section uncertainties, and possibly near and far detector differences in reconstruction and measurement of neutrino energy.  It is clear that it is a big challenge to account for all these effects and provide a small uncertainty on the far detector prediction. On top of this, far detector events of greatest interest for $\delta_{CP}$ measurement are interactions of electron neutrinos that come from oscillations, whereas the near detector event rate is dominated by un-oscillated muon neutrino interactions. \\



So far both LBNF/DUNE and T2HK have considered to have a single detector located at the edge of the experimental sites. For T2HK the near detector is foreseen to be the upgrade of the current off-axis near detector (ND280) of the T2K experiment. The location is then kept at 280~m from the JPARC production target. The upgrade of the ND280 detector is already on-going~\cite{ND280upCDR} and will serve also for the remaining data taking of T2K. New technologies, such as SuperFGD~\cite{superFGD} and resistive Micromegas~\cite{resMM}, will be used. The upgraded version of ND280 addresses some of the problems discussed in this document, in particular, a more precise reconstruction of the incoming neutrino energy and a better handle on nuclear effects thanks to lower momentum threshold for reconstructed protons. Another important feature of the new detector configuration is the possibility to cover a similar kinematic phase-space (4$\pi$ acceptance) for neutrino interactions as in the far detector. 

For LBNF/DUNE the near detector site is chosen at  574 m from the production target.
The concept of an hybrid detector has been chosen to serve several functions:
1) a liquid argon TPC provides the same target as the far detector, and a similar event reconstruction 2) a magnetized muon detector 
measures the spectrum of muons, 3) a fine grained detector measures the exclusive states with excellent resolution to provide accurate input to the detector simulations\cite{DUNENDtalk}.  
The broadband on-axis LBNF beam is very intense, enabling very high statistics 
which might allow to use neutrino electron elastic scattering to obtain the absolute neutrino flux, and the energy scale.



As mentioned above for LBNF/DUNE and T2HK, the spectrum of neutrinos as measured at the chosen near site will require significant correction to extrapolate to the far site.  
Given the need to minimise the error on the prediction of the neutrino spectrum at the far detector to accomplish the physics goals, several proposals have been made to operate multiple detectors either at different distances or at different off-axis angles for both LBNF/DUNE and T2HK.  We briefly mention them here with their advantages and the current 
choices. These proposals were discussed in the CENF-ND Forum and the presentation materials are available on the website~\cite{CENFND}.

When extrapolating the measurements in near detectors to predict the far detector event rates, it is necessary to apply the neutrino energy dependent oscillation probability to the neutrino energy dependent rate in the near detectors.  However, since the neutrino energy is not directly observed in the near detector, a deconvolution must be applied to extract the energy dependence.  This is an under-constrained problem, so prior knowledge of the neutrino flux and interaction models must be applied to find a solution.  Consequently biases in the flux and interaction models will introduce biases in the extrapolated prediction at the far detector.  

The PRISM method allows for near detector measurements at a range of off-axis angles where narrow-band neutrino energy spectra are selected according to the well understood two-body pion decay kinematics. 
This approach allows to establish a direct connection between the observable measured at the near detector and the energy of the neutrinos.  This additional information is added to the deconvolution problem with the goal of over-constraining the problem with data so that a model independent extrapolation can be made.  

This approach has been adopted in current conceptual designs for near and intermediate detectors by both the Hyper-K and DUNE experiments. The PRISM method was proposed for the first time for the NuPRISM detector (now J-PARC E61 experiment~\cite{NuPRISMproposal}) based on water Cherenkov technology. The intermediate detector site in the T2HK experiment is not defined yet, sites between 0.75~km and 2~km are being evaluated. For LBNF/DUNE the PRISM concept is being studied by considering a movable liquid argon TPC detector in the 574~meters site and thus extending the hall perpendicularly to the beam direction. \\

{\it \noindent The CENF-ND working group considers a very high priority to ensure that the near detector facilities at both LBNF/DUNE and T2HK are built in a way that an array of detectors can be accommodated; therefore, the facility will serve the experimental program in the long term as it evolves.  }

\section{Ancillary measurements}
\label{sec:AncMeas}
Dramatic improvements in LBN analyses have been achieved over the last 10 years and the next generation of experiments aim to make very precise measurements of the oscillation parameters. In parallel to the refinement of the analysis techniques, the LBN community acknowledges the compelling need of parallel measurements to the main neutrino interaction observation at the near and far detector locations.
The goal of such ancillary measurements is to address some of the current main issues affecting the oscillation analyses which need dedicated studies.
Ancillary measurements can be for example related to the detector response in terms of both calibration and reconstruction issues which, propagated to the oscillation analysis, can introduce a sizable bias in the parameter measurements. This is  the case of neutrons as detailed in Section~\ref{sec:Neutrons}.
In parallel, small-size experiments can be run to perform close studies of some physics processes such as the neutrino production from meson decay, neutrino-nucleus interactions, the re-scattering of primary neutrino interaction products (final state interaction, FSI). Those kind of measurements can run in parallel to the main LBN experiments and even on a different site. Section~\ref{sec:LowPbeamline} describes the measurements which can be performed with a very -- very low momentum beamline (below 1~GeV/c).  

A remarkable example of such kind of ancillary measurements is the NA61/SHINE experiment~\cite{NA61} at the CERN SPS which is measuring pion and kaon yields in the full phase space relevant for neutrino beams. Such measurements allow to control the uncertainties related to the neutrino beam production. The positive impact of these measurements on neutrino oscillation experiments is clearly reported by the T2K experiment~\cite{T2KFlux}.
Recent NA61/SHINE measurements performed with a T2K replica target should allow to constrain the (anti-)neutrino flux predictions at T2K
down to an unprecedented precision of about 5\%~\cite{NA61_2009data}.
Similarly, the MINER$\nu$A experiment also uses hadron production measurements from an external experiments (NA49 \cite{NA49} and MIPP~\cite{MIPP}) to achieve an \emph{a priori} flux prediction with less than 10\% uncertainty with no constraints from the detector~\cite{ppfx}. 
Additional hadron production measurements relevant for Fermilab-based neutrino beams are currently being performed by NA61/SHINE. Future measurements, in particular, with replicas of DUNE and T2HK targets are being planned. Potential NA61/SHINE measurements with incoming low momentum beam (1--10~GeV/c), which requires modifications of the H2 beam line in the North Area at CERN, could be important for further reduction of neutrino flux uncertainties in both accelerator-based and atmospheric neutrino experiments. \\


In this Section we suggest few topics which are related to the dominant uncertainties affecting the current oscillation analyses.

\subsection{Impact of Neutron reconstruction in LBN experiments }
\label{sec:Neutrons}

Neutrino interactions are susceptible to have neutrons among the final state particles. Neutron multiplicity and energy vary for neutrino and anti-neutrino events and deeply depend on the reaction mechanism. 
The missed detection of such particles has consequences in all neutrino analyses both atmospheric and accelerator-based neutrino studies as well as in the searches for proton decay or supernova relic neutrino detection (SRN). 
Therefore the possible detection of neutrons in both near and far detectors of LBN experiments would clearly increase the physics potential of such experiments, allowing a reduction of different type of backgrounds in which neutrons are present.

\begin{figure}
 \centering
 \begin{subfigure}{.45\textwidth}
   \centering
    \includegraphics[width=.9\linewidth]{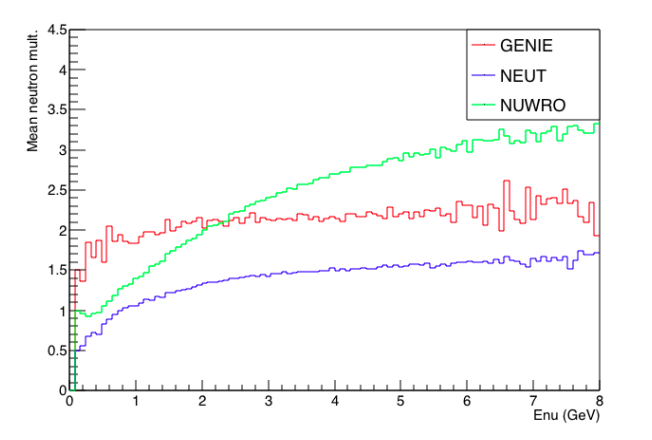}
   \caption{ \label{fig:Nmean_FHC} }
 \end{subfigure}%
 \hspace{.2cm}
 \begin{subfigure}{.45\textwidth}
   \centering
   \includegraphics[width=.9\linewidth]{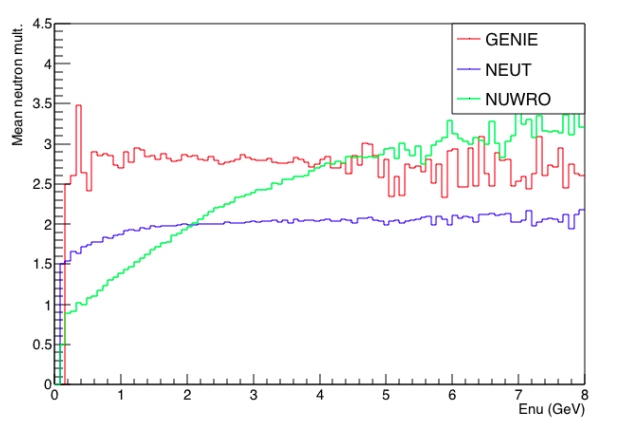}
   \caption{\label{fig:Nmean_RHC} }
   \end{subfigure}
 \caption{\label{fig:neutronsIntro} Dependence of the mean primary neutron multiplicity on the neutrino energy in a simulation for DUNE beam FHC $\nu_\mu$ CC (a) and RHC $\bar{\nu}_\mu$ CC events (b). }
 \end{figure}

\begin{figure}
 \centering
 \begin{subfigure}{.5\textwidth}
   \centering
   \includegraphics[width=.9\linewidth]{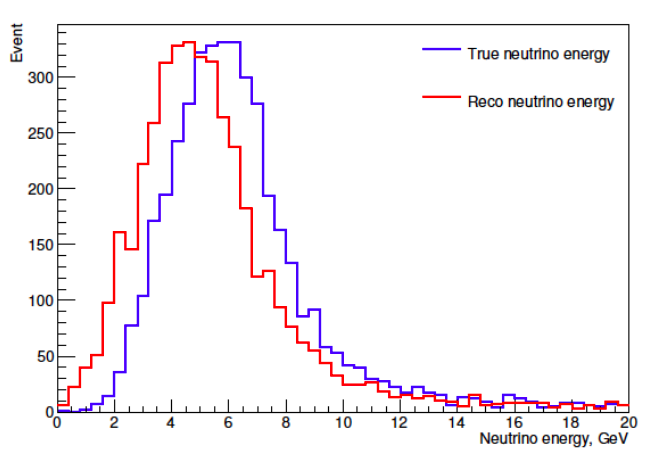}
   \caption{ \label{fig:ImpactNinEnergy} }
 \end{subfigure}%
 \begin{subfigure}{.5\textwidth}
   \centering
   \includegraphics[width=.9\linewidth]{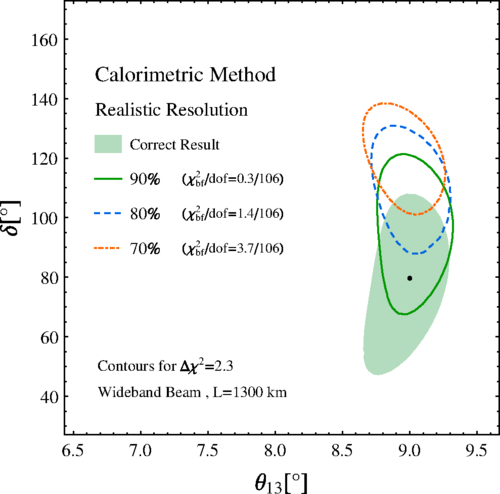}
   \caption{\label{fig:ImpactNdCP}  }
   \end{subfigure}
 \caption{\label{fig:LBNoscNimpact} (a) True neutrino energy distribution of events (blue); in the red curve, the neutrino energy is obtained assuming no neutron reconstruction and perfect reconstruction of other particles~\cite{talkCAPTAIN}. (b) Study of the possible impact of missing energy reconstruction from neutrons on the $\delta_{CP}$ parameter~\cite{MissingEnergydCP}. }
 \end{figure}

To give an example, the characteristic signature of the simplest neutrino interaction mode, Charged Current Quasi Elastic (CCQE) scattering, is the outgoing lepton and a proton (neutron) for neutrino (anti-neutrino) interactions. At  higher energies, the number of the ejected nucleons increases but a higher number of neutrons is still more common  for anti-neutrino events. 
Figure~\ref{fig:neutronsIntro} shows, as example,  the mean number of neutrons expected from various simulation of $\nu_\mu$ ($\overline{\nu}_\mu$) primary interactions as a function of the neutrino energy for DUNE\cite{NeutronImpactDUNE}. 
Neutron detection can thus give a handle to the LBN far detectors, which are usually non-magnetized detectors, to discriminate between neutrino and anti-neutrino interactions. 
At the far detectors of LBN experiments neutron detection is crucial for a  correct neutrino energy reconstruction based on calorimetry (e.g. DUNE, the SBN program). The missing contribution from neutrons can lead to biases in the reconstructed neutrino energy  (Figure~\ref{fig:ImpactNinEnergy}) and, therefore, on the measured oscillation parameters (Figure~\ref{fig:ImpactNdCP}).  

Neutron tagging can furthermore benefit proton decay analyses, allowing for a better rejection of the atmospheric neutrino background. 
The absence of a neutron in a detector able to tag these particles would improve the signal strength of a potential proton decay signature. Similar considerations also apply for the detection of SuperNovae explosions with neutrinos.

Finally, the multiplicity and energy of neutrons from neutrino and anti-neutrino interactions is not fully understood today since severe discrepancies are observed among different neutrino generators. 
This is clearly shown by Figure~\ref{fig:neutronsIntro} where three neutrino event generators, NEUT~\cite{NEUT}, GENIE~\cite{GENIE} and NUWRO~\cite{Nuwro} are compared. 
Furthermore, efficient neutron detection can help to better constrain nuclear effects such as two-particle-two-hole (2p2h) excitations. 


Following all the considerations above, dedicated measurements to understand and characterise detector responses to neutrons is highly recommended to prepare the next generation of LBN experiments. The use of tagged neutron beams instead of radioactive sources would allow measurements to cover a large energy range and better control the rate and the direction of the particles. A neutron beam line already in use also for this kind of purposes is available at Los Alamos National Laboratory. A second beam line, as for example the one for the nTOF experiment at CERN, could be of deep interest for the characterisation of the components of the Near detector facilities foreseen for both T2HK and DUNE. This is even more true if one considers the deep implication of European Institutions in these detectors.  

\subsection{Low Momentum beam line }
\label{sec:LowPbeamline}

Present and next generation of accelerator-based neutrino experiments deal with neutrino energies in the range 0.1-10~GeV.   Secondary particles produced in these interactions are concentrated in the sub-GeV to few-GeV energy range. The need to expose detector prototypes to charged particle beams in this energy range is therefore crucial. Efforts in this direction have already been carried out for example at Fermilab with the LArIAT testbeam, and at CERN with the low energy beamlines for the ProtoDUNE detectors. However, the very low momentum  range, below 1~GeV/c, remains vastly unexplored. Data is needed for calibrations, for tests of reconstruction, for Monte Carlo benchmarking. Strong interest for a low-energy beamline has been expressed by many groups involved in  different detectors technologies, from water Cherenkov to scintillators and Liquid Argon, etc. At the same time, the successful operation of the H4-VLE beamline for ProtoDUNE has demonstrated that low energy beams can be produced at CERN, and can be instrumented with performant beam monitors, momentum spectrometer  and Time-Of-Flight devices without degrading the beam. H4-VLE was operated down to 0.5~GeV momentum, where, however, almost only electrons were surviving to the relatively long flight. More optimization is surely needed to reach the sub-GeV goal. However, having the "very-very low energy beamline" operational after the long shutdown would be an important contribution to all neutrino projects around the world, independent on the level of completion, operation or design.  

Low energy test beams are of crucial importance in order to tune the pion and proton re-scattering inside (FSI) and outside (Secondary Interactions) the nucleus where the neutrino interaction happens. Such effects must be precisely known in order to disentangle the different neutrino interaction channels characterized by different pion and proton multiplicities. The cross-sections of different processes have, indeed, different energy dependence thus they need to be measured separately at the ND and extrapolated separately at the far detector, which has a different neutrino energy spectrum because of oscillations. The knowledge of FSI and SI is also crucial for the correct reconstruction of neutrino energy both in LAr and Water Cherenkov detectors.
Measurements with electron scattering data have shown very large effects: e.g. the probability for a proton to leave the nucleus without re-interacting after the main neutrino interactions is only 60\% for Carbon~\cite{Adependence}, and it will be even less for argon. Similarly, neutrino scattering models predict that about 50\% of the pions reinteract inside the Carbon nucleus. Such models are tuned on the basis of pion and proton scattering data at very low energy.
Another important possibility to explore is the usage of low energy beams for new electron scattering measurement which provide a crucial input to the neutrino interaction modeling, as will be discussed in the next section.

The available data on both for pion and proton scattering, relevant for this aim, are sparse. Old electron scattering data are available but the model corrections needed to extract the relevant quantities from the data are often outdated and/or the reported measurements miss the relevant information to exploit them to the needed precision (e.g. missing correlations in the uncertainties). Two recent and notable exception are the DUET measurement of pion scattering at TRIUMF ~\cite{DUET} and the E12-14-012 electron-scattering measurement at JLab~\cite{JLAB, JLAB1, Benhar}. Both are being extensively exploited by the LBN community for the improvement of recent and future oscillation measurements.\\

{\it \noindent A systematic campaign of such electron, pion and proton scattering in beams at low energy (sub-GeV) can be performed at CERN and, while running 'parassitically' to the program of new detector testing and calibration, it would be of invaluable benefit to reach the precision goals of the next generation LBN experiments.}

\section{ A Forum for Theory and Neutrino Event Generators}
\label{sec:Generators}


Uncertainties associated with neutrino interaction cross-section models need to be reduced down to the few percent level to match the experimental requirements. Since the experiments rely on the interactions
of neutrinos with bound nucleons inside atomic nuclei, a good understanding and realistic modeling  of the hadronic and nuclear physics of these interactions is mandatory. A detailed account of the present status and open questions in these studies are given in Ref.~\cite{WhitepaperNUSTEC}. The availability of new data from running neutrino experiments has stimulated a considerable theoretical activity, leading to a better description of the relevant processes. Nonetheless, the level of precision is still far from the desired one.

Many different neutrino event generators are presently available, such as  GENIE \cite{GENIE}, NEUT~\cite{NEUT}, NuWro~\cite{Nuwro}, GiBUU~\cite{Gibuu, Gibuu2} and NUNDIS (in FLUKA)\cite{Nundis, FLUKA}. Some of these generators are not well documented neither in their physics content nor in their numerical realization. It will, therefore, be a challenge for the coming years to improve on this situation and narrow the generators down to those based on the best hadronic and nuclear physics available. Generators are an essential ingredient of any experiment and have to be as much state-of-the-art as the experimental equipment. They should thus also undergo the same level of scrutiny as the experiments. 

The development and validation of reliable theoretical descriptions and generators are complicated by many different factors.  Precise theoretical knowledge and description of neutrino-nucleon interactions is not yet granted~\cite{talkARuso}. The lack of precise data for neutrino scattering on Hydrogen and Deuterium targets limits our ability to constrain relevant quantities like the nucleon and nucleon-to-resonance transition form factors. This deficiency might be partially compensated in the future by 
lattice QCD simulations. 
The extension of the energy range
demands a consistent description of the transition region between different interaction channels, and is a challenge for models that have to be pushed to their limit of validity,
as discussed in the recent NUSTEC workshop on DIS~\cite{DISworkshop}. 

\begin{figure}[tbh]
 \centering
 \begin{subfigure}{.47\textwidth}
   \centering
   \includegraphics[width=.95\linewidth]{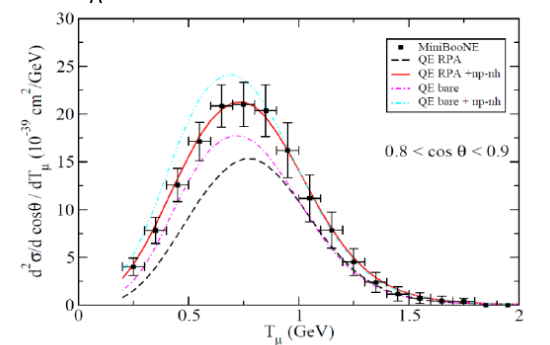}
    \caption{\label{fig:Miniboone_Martini}}
 \end{subfigure}
 \hspace{.5cm}
 \begin{subfigure}{.47\textwidth}
   \centering
   \includegraphics[width=.95\linewidth]{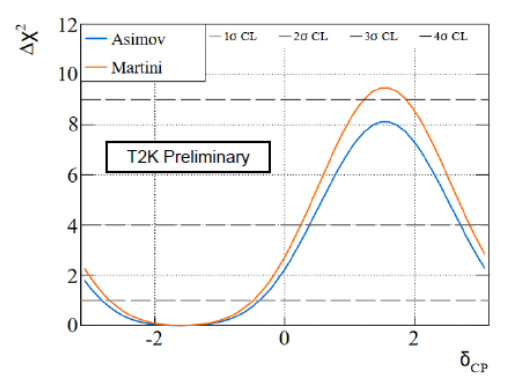}
   \caption{\label{fig:Sensitivity_Martini} }
   \end{subfigure}
    \caption{(a) MiniBooNE flux-averaged CCQE $\nu_\mu - ^{12}$C double differential cross section per neutron for $ 0.8 < \cos{\theta} <0.9 $ as a function of the muon kinetic energy. Lines show the contribution of different mechanisms to the cross section according to Martini {\it et al.} \cite{Martini}. (b) Preliminary study of the T2K sensitivity with the standard T2K Monte Carlo and with the Martini 2p2h model. From \protect\cite{Scott}.}
  \end{figure}

To add an order of magnitude of complexity, the nuclear environment implies initial state and final state effects which modify the cross sections and change the event topology, mixing different primary reaction channels. This scenario is apparent in the fact that the MiniBooNE quasilastic scattering data actually contained a sizable fraction of events from neutrino interactions on nucleon pairs (Figure~\ref{fig:Miniboone_Martini}). This is a source of bias in kinematic neutrino energy reconstruction and propagates to the experimental sensitivity for the determination of neutrino properties as in Figure~\ref{fig:Sensitivity_Martini}. 

Comparisons with data themselves are hindered by the fact that in wide-band neutrino beams the initial neutrino energy is not known and has to be inferred from final state quantities. Some data sets are model dependent, and sometimes contradictory; see, for example, the inconsistency between Minerva and MiniBooNE data on single pion production pointed out in \cite{mbpions}.  A better connection with experiments and experts in electron-scattering data would be an asset to compare models with cleaner and more precise data.

 To perform their task in both cross-section and oscillation measurements, 
 event generators require a flexible structure to accommodate different models, adequate implementation of the steering parameters and most relevantly, validity across a wide energy range. These generators should be tested against the most precise electron scattering data.  
 The existence of a common generator used across experiments is, at the same time, a key element for future combined results from experiments aiming at exploiting the complementary between them.  Contrary to other fields in particle physics, the neutrino generators have been developed within particular experiments with some remarkable exceptions. Adopting the organisational scheme from LHC generators may be of great value for the community, allowing theorists to contribute more actively  and facilitating the integration of newly developed models. Europe has been at the forefront with generators such as GENIE, NuWro, GiBUU and FLUKA. This fact, together with the presence of very strong experimental and theoretical groups, positions Europe in an optimal situation to lead this development. Several initiatives have begun, such as two workshops at the ECT* in Trento~\cite{TRworkshop}. The promotion of these activities at CERN and at the different European Research Institutions are critical to the success of the neutrino oscillation programme and to maintain Europe in a strong leadership position.
 
 These actions should go beyond the organisation of workshops and should include the active promotion of the activities and the creation of a career path for the researchers involved in the modeling of neutrino interactions and generator improvement. The expertise needed to develop neutrino interaction theories and generators must come from several different communities from low energy nuclear physics to particle physics. \\

{\it \noindent A CERN-based forum collecting ideas,  providing opportunities to work together on timescales longer than standard workshops, forming young scientists, would surely enhance the possibility to collect the most from all different fields.}

\section{Conclusions}
\label{sec:Conclusions}

European physicists using the facilities at CERN and other European laboratories can play a leading role in the program of precision neutrino physics with the long term focus on sub-percent level of
understanding of accelerator neutrino beams and of neutrino interactions in the near and far detectors. Our recommendations are:

\begin{itemize}
\item Supporting a strong involvement of European groups in the design, construction, and upgrade of Near Detectors
    \item Supporting the excavation of near detector caverns large enough to allow for multiple detectors, or movable detectors such as nuPRISM,  and future upgrades
    \item Supporting ongoing and future experiments measuring hadron production at energies/targets of interest (such as NA61/SHINE at CERN)
    \item Supporting new ideas for controlled neutrino cross section measurements, such as the ENUBET project
    \item Providing facilities where calibrated neutron beams are available. For examples, setting up a test-beam area at the nToF facility
    \item Providing low energy (subGeV) hadron and electron beams, for detector calibration and cross section measurements. 
    \item Fostering theoretical studies and improvements of Monte Carlo generators of neutrino interactions, for instance with the creation of a neutrino centre providing short and long term work opportunities, computing facilities, workshops, schools, etc.

\end{itemize}


\bibliographystyle{unsrt}
\bibliography{sample}

\end{document}